\documentclass[reprint, aps, pra, twoside, floatfix, longbibliography, superscriptaddress]{revtex4-2}

\usepackage{amsmath, amssymb}
\usepackage{graphicx}

\usepackage{hyperref}
\hypersetup{hidelinks}

\usepackage{orcidlink}

\newcommand{\bra}[1]{\langle{#1}|}
\newcommand{\ket}[1]{|{#1}\rangle}
\newcommand{\Ham}{\mathcal{H}}			
\newcommand{\res}{\mathrm{r}}  			
\newcommand{\atom}{\mathrm{a}}			
\newcommand{\cpl}{\mathrm{c}}			
\newcommand{\JJ}{\mathrm{J}}			
\newcommand{\ext}{\mathrm{e}}			
\newcommand{\mi}{\mathrm{min}}			
\newcommand{\zpf}{\mathrm{zpf}}			
\newcommand{\PhiJ}{\Phi_\JJ}	
\newcommand{\Phie}{\Phi_\ext}	
\newcommand{\Phic}{\Phi_\cpl}	

\begin{document}

\title{Two-photon coupling via Josephson element I:
\\
Breaking the symmetry with magnetic fields}

\author{E.~V.~Stolyarov\,\orcidlink{0000-0001-7531-5953}}
\affiliation{Quantum Optics and Quantum Information Group, Bogolyubov Institute for Theoretical Physics, National Academy of Sciences of Ukraine, vul.\ Metrolohichna 14-b, Kyiv 03143, Ukraine}
\author{V.~L.~Andriichuk\,\orcidlink{0000-0001-6004-7175}}
\affiliation{Institute of Physics of the National Academy of Sciences, pr.\ Nauky 46, Kyiv 03028, Ukraine}
\author{A.~M.~Sokolov\,\orcidlink{0000-0002-2691-317X}}
\email[E-mail: ]{andriy145@gmail.com}
\affiliation{Institute of Physics of the National Academy of Sciences, pr.\ Nauky 46, Kyiv 03028, Ukraine}

\begin{abstract}
We consider a coupling element based on a symmetric superconducting quantum interference device~(SQUID) and show that it mediates a two-photon interaction.
This and other inductive interactions due to the SQUID can be switched off \emph{in situ}.
We derive the system Hamiltonian for coupled resonator and rf SQUID.
The rf SQUID dwells in the vicinity of its metastable well holding a number of  energy states and acts as an artificial atom.
We discuss how the Josephson symmetry breaks owing to magnetic fields in the superconducting loops---both directly and via a non-zero superconducting phase difference over the rf SQUID.
We assess that the two-photon coupling strength reaches 18\,MHz which can exceed the single-photon capacitive interaction in the coupler.
\end{abstract}

\maketitle

\section{Introduction} \label{secIntroduction}

A nonlinearity cubic in the system variables is the first one that can appear in the coupling energy.
In the equations of motion, it spawns a quadratic interaction term;
typically, only the linear interaction exceeds this coupling strength.
Naturally, this nonlinearity is widely used in applications, such as lasing~\cite{louisell1964radiation}, radio-frequency homo- and heterodyne detection~\cite{blais2021circuit}, optomechanical cooling and amplification~\cite{aspelmeyer2014optomechanics}, parametric amplification~\cite{louisell1964radiation,esposito2021perspective,anisimov2003kolyvannya1} and frequency conversion~\cite{louisell1964radiation,zakkabajjani2011quantum,kurizki2015quantum,xiang2013hybrid,lu2023resolvingfock}, to name a few.

For interacting radiation mode and matter particle, this nonlinearity can give rise to two-photon processes---when the coupling energy is quadratic in the mode variables.
In that case, two photons excite the particle one step up in its energy levels, while a step down releases two photons back into the radiation mode.
Normally, this nonlinearity is small compared to the linear coupling and such processes only become visible for higher field strengths~\cite{nakamura2001rabi,lisenfeld2007thesis,valimaa2022multiphoton}.
Perturbative two-photon coupling can be arranged via non-resonant single-photon processes as in Refs.~\cite{ashhab2006rabi,garziano2015multiphoton,sokolov2020superconducting,sokolov2023thesis}.

\begin{figure}
\includegraphics{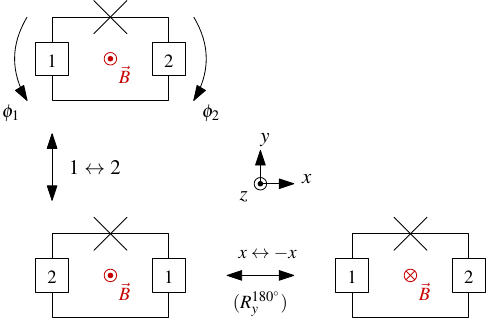}
\caption{Exchanging the elements 1 and 2 in the superconducting loop makes the circuit different if a magnetic field $\vec B$ is present.
That is, inversion $x \leftrightarrow -x$ (or $180^\circ$ rotation around the $y$ axis) does not bring the circuit into the configuration before the exchange of the elements.
In other words, the system is potentially \emph{not} invariant with respect to the exchange of the elements.
Hence a term can be present in the Josephson energy that is odd in the drops $\phi_1$ and $\phi_2$ of the superconducting condensate phase.
}
\label{figSymmetry}
\end{figure}

We focus on \emph{non-perturbative} two-photon interaction via the Josephson nonlinearity.
We assume that a coupling linear in a given system variable induces a single transition, while a quadratic one induces two transitions~\footnote{
Note that this is not always the case.
In Ref.~\cite{zou2020multiphoton}, it was proposed to mediate the two-photon interaction via modulation of the Josephson energy of a Cooper pair box~\cite{bouchiat1998quantum,devoret1997} qubit.
There, a single transition in the charge basis of the qubit arises due to the coupling energy quadratic in the qubit phase variable.
In that case, a nonlinearity quartic in the system variables yields two-photon coupling.
}.
However, in the absence of magnetic fluxes, Josephson coupling energy is symmetric in the superconducting phase difference $\phi_1 - \phi_2$.
It is normally~\cite{golubev2004currentphasejj,heikkila2000nonequilibrium,heikkila2000supercurrent} expressed as~\cite{devoret1997,vool2017,rasmussen2021sccompanion}
\begin{equation}
\label{eqJosephsonEnergyNoField}
	U_\cpl^{\vec B = 0} \propto \cos(\phi_1 - \phi_2).
\end{equation}
This symmetry prevents cubic contributions with $\phi_1^2 \phi_2$ and $\phi_1 \phi_2^2$ in the energy and hence two-photon processes are absent~\footnote{For other examples of symmetries precluding a related interaction with cubic energy, see Ref.~\cite{sokolov2024signatures} and the references therein.}.
Mathematically, an additional static phase should enter the cosine argument to break the Josephson symmetry and allow the cubic terms in the expansion.
Then the two-photon processes may be observed:
Supposing that one system holds photonic excitations, their frequency should be roughly twice as large as in the other, matter-like system.

In one possibility, Josephson symmetry is broken by pumping one of the system parts.
A steady state can be then created, such as the quartic energy terms become cubic in small departures from it~\cite{lescanne2020irreversible,balembois2024cyclically}.
In the other approach, the symmetry is broken with a magnetic field as explained in Fig.~\ref{figSymmetry}.
This approach is often used to achieve three-wave mixing in Josephson parametric amplifiers~\cite{abdo2021hifi,zorin2016jtwpa3wave,frattini2017threewave,khabipov2022noncentrosymmetric} and in related proposals~\cite{felicetti2018twophoton,felicetti2018ultrastrong,khabipov2022noncentrosymmetric} for two-photon interaction in superconducting circuits.

In our work, we explore the symmetry breaking directly induced by the external magnetic field, as well as a related mechanism.
The latter relies on the equilibrium phase differences at the coupled elements, $\phi_1^\mi$ and $\phi_2^\mi$, being non-zero in a metastable state.
Then, if $\phi_1^\mi \ne \phi_2^\mi$, the Josephson symmetry breaks in the coupling energy in Eq.~\eqref{eqJosephsonEnergyNoField}.
A cubic energy term may arise in the small departures $\phi_{1,2} - \phi_{1,2}^\mi$ from the equilibrium.
Similarly to Ref.~\cite{sokolov2020superconducting}, we are interested in such contribution that is quadratic in one of the phases---the phase of a resonator.
Another coupled system, an rf SQUID~(see Fig.~\ref{figScheme}) with an asymmetry in its ground state, then provides the required equilibrium phase shift.
The systems interact via a \emph{separate} Josephson coupling element.

Indeed another difference with Refs.~\cite{felicetti2018twophoton,felicetti2018ultrastrong,zou2020multiphoton,sokolov2020superconducting} is that we separate the coupling nonlinearity.
That is similar to Ref.~\cite{ayyash2024driven} published recently and Ref.~\cite{khabipov2022noncentrosymmetric};
although, these works do not consider symmetry breaking due to non-zero equilibrium phases.
Separate coupling may ease connecting other types of superconducting circuits or even systems of a different nature~\cite{xiang2013hybrid, kurizki2015quantum}.
Moreover, with a coupling dc SQUID as in Fig.~\ref{figScheme}, one can switch off all inductive interactions, by biasing it appropriately with external magnetic field.
This very bias can yield an additional symmetry-breaking shift in the phases, due to a mechanism similar to that in Fig.~\ref{figSymmetry} and in Refs.~\cite{abdo2021hifi,zorin2016jtwpa3wave,frattini2017threewave,felicetti2018twophoton,felicetti2018ultrastrong,ayyash2024driven}.
We estimate the two-photon coupling while taking into account both of the shifts.

The paper is organized as follows.
We focus on the rf SQUID circuit acting as an artificial atom and provide the full circuit Hamiltonian in Sec.~\ref{secModel}.
First we write out the system classical Hamiltonian in Sec.~\ref{secClassicalHamiltonian} and discuss our use of the flux quantization rule in Sec.~\ref{secCurrentFreePath}.
The classical Hamiltonian allows us to observe in Sec.~\ref{secBiasAndPhiMins} that the coupler magnetic bias can be seen as shifting the equilibrium phases of the coupled systems.
We further develop some intuition about the system dynamics, identify the suitable regime, and find the equilibrium phases in Sec.~\ref{secEquilibrium}.
We then quantize the system in Sec.~\ref{secQuantumHamiltonian} around its equilibrium point.
In Sec.~\ref{secTwoPhoton}, we estimate the two-photon coupling strength;
then, in Sec.~\ref{secDiscussion} we discuss approximations, assumptions, and limitations in our treatment.
We provide some possible applications in Sec.~\ref{secApplications} and conclude in Sec.~\ref{secConclusions}.
Detailed derivation of the circuit Hamiltonian is given in the \hyperref[apCircuitHamiltonian]{Appendix}.

\section{Model}
\label{secModel}

\begin{figure}
\centering
\includegraphics{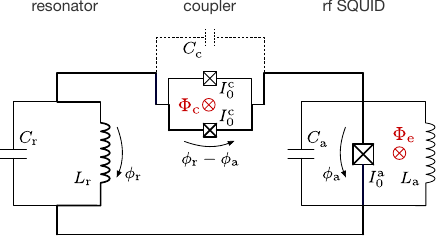}
\caption{
A resonator connects to an artificial atom via the SQUID coupling.
An rf SQUID can act as an artificial atom.
The coupling SQUID may be shunted with a capacitance.
External fluxes in the circuit allow for engineering of the two-photon interaction.
The \emph{big loop}~(thick) contains the atom junction, the coupler bottom junction, and the resonator inductance.
As discussed in Secs.~\ref{secCurrentFreePath} and \ref{secDiscussion}, we assume that there is a path with negligible current and resistance inside the superconducting loop.
}
\label{figScheme}
\end{figure}

We consider a microwave resonator that is nonlinearly coupled to an artificial atom.
To be concrete, we consider a capacitively-shunted rf SQUID acting as an artificial atom.
The schematics of the setup is illustrated in Fig.~\ref{figScheme}.

To simplify the theory, we consider a single-mode resonator implemented by a lumped-element $LC$ oscillator \cite{allman2010, collodo2019, vrajitoarea2020} with inductance $L_\res$ and capacitance $C_\res$.
The rf SQUID is a superconducting loop of inductance $L_\atom$ that is interrupted by a Josephson junction with a critical current $I^\atom_0$ and a self-capacitance $C^\atom_\JJ$.
In addition to the junction self-capacitance, we consider a shunting capacitance  $C_\atom$ in parallel~\cite{opremcak2018measurement,opremcak2020high} with it.
An external magnetic flux $\Phie$ threads through the loop.
Such a configuration is similar to the phase qubit in Ref.~\cite{bennett2009decoherencerfsquid} or Josephson photomultipliers in Refs.~\cite{opremcak2018measurement,opremcak2020high,shnyrkov2023rfsquid,stolyarov2023pnr,ilinskaya2024fluxqubit}.

The resonator and the atom interact through a nonlinear coupler---a dc SQUID~\cite{neumeier2013single, collodo2019}.
It consists of two identical Josephson junctions, with critical currents $I^\cpl_0$ and self-capacitances $C_{1\JJ}^\cpl$.
It may be shunted with an additional capacitance $C_\cpl$.
Each junction energy is $E^\cpl_{1\JJ} = I^\cpl_0 \Phi_0/2\pi$ with $ \Phi_0 = h/2e$ being the flux quantum.
The coupling SQUID is biased by the external flux $\Phic$.
A related scheme with a coupling SQUID was theoretically studied~\cite{geller2015tunable} and used~\cite{chen2014tunable} for arranging a tunable single-excitation coupling between transmons.

\subsection{Classical Hamiltonian} \label{secClassicalHamiltonian}

The Hamiltonian of the setup in Fig.~\ref{figScheme} consists of the resonator, atom, and the coupling parts:
\begin{equation} \label{eqClassicalHamiltonian}
	\Ham = \Ham_\res + \Ham_\atom + \Ham_\cpl.
\end{equation}
We provide its detailed derivation in the \hyperref[apCircuitHamiltonian]{Appendix}.
Here we provide the expressions required for the further treatment and highlight the physics that is important for us.

The first term in Eq.~\eqref{eqClassicalHamiltonian} describes a single-mode resonator and reads as
\begin{equation} \label{eq:ham_res}
	\Ham_\res = 4 E^\res_C n^2_\res + U_\res, \quad U_\res = \frac{E^\res_L}{2} \phi_\res^2.
\end{equation}
Here $U_\res$ denotes the resonator potential energy, while $E^\res_C = e^2/2\tilde{C}_\res$ and $E^\res_L = \Phi_0^2/4\pi^2L_\res$ are its charging and the inductive energies.
Tilded capacitances, such as $\tilde C_\res$, denote their values as renormalized by the capacitive coupling.
We provide precise expressions for them in the \hyperref[apCircuitHamiltonian]{Appendix};
under most circumstances, the capacitive coupling is small and $\tilde C_\res \approx C_\res$ etc.
In a lumped-element resonator that we have, $n_\res$ is the charge at the capacitor, measured in the units of the Cooper pair charge and $\phi_\res$ is the drop of the superconducting condensate phase across the capacitor.

As we only have regular junctions~\footnote{Current through the regular junction with a phase difference $\phi$ is $I_0 \sin\phi$, where the critical current $I_0 > 0$, as opposed, for example, to the $\pi$ junctions~\cite{golubev2004currentphasejj,heikkila2000nonequilibrium,heikkila2000supercurrent}.} in the circuit, junction phase differences are zero in equilibrium with no magnetic fields $\{\vec B_i\}$ applied.
Moreover, we do not restrict~\cite{barone1982physicsappsjj,rasmussen2021sccompanion} the phase to the region from 0 to $2\pi$, unlike Refs.~\cite{devoret1997,vool2017}.
Then $\phi_\res$ can be expressed in terms of the respective voltage drop $V_\res$:
\begin{equation}
\label{eqNodeFlux}
	\phi_\res = \frac{2\pi}{\Phi_0}
		\int_{\substack{t'=-\infty \\ \{\vec B_i = 0\}}}^t dt' V_\res,
\end{equation}
which is equal, up to the choice of flux units, to the definition of the node flux as in Refs.~\cite{devoret1997,vool2017,rasmussen2021sccompanion}.
In our case, equilibrium phase differences \emph{with} magnetic fields are of interest.
Then, the fields are adiabatically turned on~\cite{devoret1997,vool2017} with $t'$ in Eq.~\eqref{eqNodeFlux}, yielding the equilibrium $\phi_\res^\mi$ of interest.
The system further evolves around such equilibrium phase differences.

The second term in Eq.~\eqref{eqClassicalHamiltonian} represents the Hamiltonian of the artificial atom, which reads as
\begin{equation}  \label{eq:ham_pd}
 \Ham_\atom = 4 E^\atom_C n^2_\atom + U_\atom,
\end{equation}
where $U_\atom$ is the atom potential energy:
\begin{equation} \label{eqAtomPotential}
  U_\atom = \frac{E^\atom_L}{2}
		   \left(\phi_\atom - \frac{2\pi\Phie}{\Phi_0}\right)^2
		   - E^\atom_\JJ \cos\phi_\atom.
\end{equation}
In the Hamiltonian in Eq.~\eqref{eq:ham_pd}, $n_\atom$ and $\phi_\atom$ are the Cooper-pair number and the phase variables of the atom, $E^\atom_C = e^2/2\tilde{C}_\atom$ and $E^\atom_L = \Phi_0^2/4\pi^2L_\atom$ stand for the atom charging and the inductive energies, and $E^\atom_\JJ$ denotes the Josephson energy of the atom.
In the case of a single junction in the artificial atom, shown in Fig.~\ref{figScheme}, the Josephson energy is $E^\atom_\JJ = E^\atom_{1\JJ} = I^\atom_0 \Phi_0 / 2\pi$.
In the \hyperref[apCircuitHamiltonian]{Appendix}, we also consider the double-junction atom whose Josephson energy can be tuned by an additional flux bias.

The last term in Eq.~\eqref{eqClassicalHamiltonian} is the coupling energy between the resonator and the atom; it reads
\begin{equation} \label{eq:ham_cpl}
	\Ham_\cpl = E^\cpl_C n_\atom n_\res + U_\cpl.
\end{equation}
The first term here describes the capacitive coupling with energy
\begin{equation}
\label{eqCouplerCapacitiveEnergy}
	E^\cpl_C = 4e^2 \frac{\tilde C_\cpl}{C'_\atom C_\res},
\end{equation}
where $C'_\atom = C_\atom + C^\atom_\JJ$ denotes the full capacitance of the atom.
The second term in Eq.~\eqref{eq:ham_cpl} is the coupler potential energy of Josephson origin:
\begin{equation} \label{eqCouplerPotential}
	U_\cpl = -E^\cpl_\JJ
			 \cos\left(\phi_\res - \phi_\atom 
                       - \frac{\pi\Phic}{\Phi_0}\right).
\end{equation}
Here, we call
\begin{equation}
\label{eqCouplerJosephsonEnergy}
	E^\cpl_\JJ = 2E^\cpl_{1\JJ} \cos(\pi\Phic/\Phi_0)
\end{equation}
the Josephson energy of the coupler, assuming that the coupling SQUID is biased so that $E^\cpl_\JJ \geq 0$.

The argument in the Josephson energy term given by Eq.~\eqref{eqCouplerPotential} is the phase drop across the coupler.
Its form is crucial for the two-photon interaction to arise.
It follows from the flux quantization rule~\cite{barone1982physicsappsjj,orlando1999threejjqubit}.
The rule states that the sum of the phase drops across the elements in the loop equals its external magnetic flux in the relevant units.
That can be shown~\cite{feynman1965feynmanlectures,berman2009measurement,barone1982physicsappsjj} by integrating the condensate phase by a path deep inside the superconducting leads, where the supercurrent density vanishes.
At the same time, we neglect the contribution in small junction gaps as in Refs.~\cite{feynman1965feynmanlectures,berman2009measurement};
alternatively, one can use the gauge-invariant phase differences~\cite{barone1982physicsappsjj,tinkham2004intro}.

\subsection{Existence of current-free path}
\label{secCurrentFreePath}

Strictly speaking, our model is always valid only when a supercurrent-free path exists in the loop, which may not be the case for type-II superconductors.
For superconducting quantum circuits, aluminum or niobium thin films are typically used~\cite{murray2021materials}.
Although aluminum is of type I in the bulk limit, aluminum films exhibit type-II superconductivity when sufficiently thin~\cite{nsanzineza2014trapping, nunez2023magnetic}.
As for niobium, it is of type II in thin films given its large penetration depth~\cite{gubin2005dependence}, although it is claimed~\cite{prozorov2022niobium} to be of type I in the bulk clean limit.

Above a certain threshold, magnetic field enters a type-II superconductor forming normal-metal spots with vortices of screening currents around them~\cite{tinkham2004intro}.
Then, if there is no current-free path due to the vortex currents, the \emph{fluxoid} quantization rule is to be used~\cite{barone1982physicsappsjj,tinkham1964consequences,tinkham2004intro}, which takes into account the currents.
Current distribution then depends on both the sample form and the field magnitude.
Hence, dependence on the external flux $\Phi_\cpl$ in Eq.~\eqref{eqCouplerPotential} can be different in that case.
It can be precisely determined with calculations as in Ref.~\cite{baelus2000vortex}.

However, we expect the \emph{flux} quantization rule to yield satisfactory results in a practically relevant case.
Far away from a vortex, the field~\cite{tinkham2004intro} and hence the screening currents decay exponentially on distances of order of the penetration depth $\lambda$.
When the system sizes and distances between vortices exceed several $\lambda$, a nearly current-free path should exist in the superconductor.
Far below the critical temperature, $\lambda$ is of several hundred nanometers for the aluminum and niobium films as thin as tens of nanometers~\cite{nunez2023magnetic,gubin2005dependence} and decreases for thicker films.
Diameters and widths of superconducting loops can easily be much larger than that~\cite{bennett2009decoherencerfsquid,orlando1999threejjqubit,stan2004critical}.

Moreover, for aluminum, the vortex-induced effects are minimized if the film is at least 155\,nm thick, which makes it mostly~\cite{stan2004critical} a type-I superconductor~\cite{nunez2023magnetic}.
It is also known, that vortices can be expelled by narrowing the superconductor width~\cite{stan2004critical,bai2025fluxtrapping}.

\subsection{Coupler bias and equilibrium phases}
\label{secBiasAndPhiMins}

The coupler bias $\Phi_\cpl$ appears directly in its Josephson energy in Eq.~\eqref{eqCouplerPotential}.
Essentially, the dc screening current invokes this phase according to the first Josephson relation---as written out in what follows.
This symmetry-breaking mechanism is related to some types of the superconducting diode effect~\cite{hou2023ubiquitous}.

In other choices of the phase variables, $\Phi_\cpl$ may be absent in the coupler Josephson energy, at least explicitly.
For example, the phase variables may be chosen so that the phase drop is $\phi_\res - \phi_\atom + \pi\Phi_\cpl/2\Phi_0$ across the top junction in the coupling SQUID (see Fig.~\ref{figScheme}) and $\phi_\res - \phi_\atom - \pi\Phi_\cpl/2\Phi_0$ across the bottom one.
According to the reasoning as in the \hyperref[apCircuitHamiltonian]{Appendix}, the argument in the Josephson energy in Eq.~\eqref{eqCouplerPotential} is simply $\phi_\res - \phi_\atom$.
The phase drop across the atom junction is $\phi_\atom + \pi\Phi_\cpl/2\Phi_0$ owing to the flux quantization condition.
In that case the coupler bias shifts the atom equilibrium phase;
it appears in the coupling Josephson energy indirectly.

Both choices of the phase variables are equivalent.
The bias can break the Josephson symmetry and facilitate its cubic nonlinearity---in the small deviations from the equilibrium.

\subsection{Point of equilibrium}
\label{secEquilibrium}

For the further analysis, we use that the system operates in the regime of dominating inductive and Josephson energies:
\begin{equation}
\label{eqPhaseIsGoodQuantumNumber}
	E^\res_L \ggg E^\res_C,
\quad
	E^\atom_\JJ \ggg E^\atom_C.
\end{equation}
From the mechanical standpoint, this resembles the motion of a heavy particle in the vicinity of its equilibrium.
Thus, we can expand the coupler potential energy in Eq.~\eqref{eqCouplerPotential} near the points $\phi^\mi_\res$ and $\phi^\mi_\atom$ of the local minimum.
These points are determined by two transcendental equations:
\begin{subequations} \label{eq:mins}
 \begin{gather}
\label{eqResonatorExtremum}
	E^\res_L \phi^\mi_\res + E^\cpl_\JJ \sin \delta = 0,
\\
\label{eqAtomExtremum}
	E^\atom_L \Big(\phi^\mi_\atom
					 - \frac{2\pi\Phie}{\Phi_0}\Big)
	+ E^\atom_\JJ \sin\phi^\mi_\atom - E^\cpl_\JJ \sin \delta
	= 0,
 \end{gather}
\end{subequations}
where the equilibrium phase difference at the coupler is
\begin{equation} \label{eqCouplerPhase}
	\delta = \phi^\mi_\res - \phi^\mi_\atom - \frac{\pi\Phic}{\Phi_0}
\end{equation}
Up to a factor, Eq.~\eqref{eqResonatorExtremum} expresses the conservation of charge that flows in and out of the resonator node with the static currents.
Equation~\eqref{eqAtomExtremum} expresses the same for the atom node.

The second term in Eq.~\eqref{eqResonatorExtremum} is proportional to the dc screening current in the big loop in Fig.~\ref{figScheme}.
In fact, Eq.~\eqref{eqResonatorExtremum} accepts another interpretation.
It follows from the flux quantization rule that the full flux threading the loop in the direction to the reader equals (see Fig.~\ref{figScheme}) $\Phi_\Sigma = \phi_\res \Phi_0 / 2\pi$.
As there is no external flux in this loop, the flux is generated entirely by the screening current $I_\mathrm{scr} = I_0^\cpl \sin\delta$ that flows clockwise.
That is,
\begin{equation}
\label{eqBigLoopFullFlux}
	\Phi_\Sigma = -L_\res I_\mathrm{scr}.
\end{equation}
Up to a factor, we have recovered Eq.~\eqref{eqResonatorExtremum}.
Note that we neglect the loop inductive energy other than stored in the resonator inductance.
We discuss the effect of possible additional inductance in Sec.~\ref{secDiscussion}.

In what follows, $\varphi_\res$ and $\varphi_\atom$ denote the departures from the equilibrium in the resonator and atom phases, i.e.,
\begin{equation}
\label{eqPhasesAroundEquilibrium}
	\phi_\res = \phi^\mi_\res + \varphi_\res,
\quad
	\phi_\atom = \phi^\mi_\atom + \varphi_\atom.
\end{equation}
In terms of them, the Josephson coupling energy reads
\begin{equation}
\label{eqCouplerPotentialViaDepartures}
	U_\cpl = -E^\cpl_\JJ \cos(\varphi_\res - \varphi_\atom + \delta).
\end{equation}
According to Eq.~\eqref{eqCouplerPhase}, the phase $\delta$ includes the equilibrium phases of each coupled system as well as the coupler bias.
This phase determines the order of the dominant inductive interaction, as we discuss later in this paper.

\subsection{Quantum Hamiltonian} \label{secQuantumHamiltonian}

We seek the quantum Hamiltonian that governs the circuit dynamics near its equilibrium at $\phi^\mi_\res$ and $\phi^\mi_\atom$.
That is, departures $\varphi_\res$ and $\varphi_\atom$ from the equilibrium are small in the resonator and atom phases as given by Eqs.~\eqref{eqPhasesAroundEquilibrium}.
We expand the cosine term in Eq.~\eqref{eqCouplerPotential} in $\varphi_\res$ and $\varphi_\atom$, while accounting for Eqs.~\eqref{eq:mins}.
Then, we promote the classical variables of the resonator and the artificial atom to the operators with the commutation relations $\left[\hat\varphi_\res, \hat{n}_\res\right] = \left[\hat\varphi_\atom, \hat{n}_\atom\right] = i$.

To arrive at the second-quantized Hamiltonian, we express these operators in terms of the bosonic ladder operators.
For the resonator variables, one has $\hat{n}_\res = -i n_{\res, \zpf} (a^\dag - a)$ and $\hat\varphi_\res = \varphi_{\res, \zpf} (a^\dag + a)$.
Here $a$ and $a^\dag$ are the photon annihilation and creation operators, while
\begin{equation}
\label{eqResZPF}
	n_{\res, \zpf} = (\tilde{E}^\res_L/32 E^\res_C)^{1/4},
\quad
	\varphi_{\res, \zpf} = (2E^\res_C/\tilde{E}^\res_L)^{1/4}
\end{equation}
are the magnitudes of the resonator charge and phase zero-point fluctuations.
Analogously, the atom operators $\hat{n}_\atom$ and $\hat\varphi_\atom$ are expressed as $\hat{n}_\atom = -i n_{\atom, \zpf} (b^\dag - b)$ and $\hat\varphi_\atom = \varphi_{\atom, \zpf} (b^\dag + b)$, where
\begin{equation}
\label{eqAtomZPF}
	n_{\atom, \zpf} = (\tilde{E}^\atom_L/32 E^\atom_C)^{1/4},
\quad
	\varphi_{\atom, \zpf} = (2E^\atom_C/\tilde{E}^\atom_L)^{1/4}
\end{equation}
are the magnitudes of the atom zero-point fluctuations.
Operators $b$ and $b^\dag$ annihilate and create an excitation in the artificial atom.

Above, we introduce the renormalized inductive-like energies of the resonator and the atom:
\begin{subequations} \label{eqInductiveEnergiesRenorm}
\begin{gather}
\label{eqRenormELr}
	\tilde{E}^\res_L = E^\res_L + E^\cpl_\JJ \cos\delta,
\\
\label{eqRenormELa}
	\tilde{E}^\atom_L = E^\atom_L + E^\atom_\JJ \cos\phi^\mi_\atom
						+ E^\cpl_\JJ \cos\delta.
\end{gather}
\end{subequations}
The last term in the equations allows for a circuit interpretation.
It describes how the atom and resonator inductances change when loaded with the Josephson inductance $\Phi_0^2/4 \pi^2 E^\cpl_\JJ$ of the coupler.

What is more important for us, the SQUID coupler---as a nonlinear inductive element---gives rise to the dipolar and various non-dipolar interactions.
In the next section, we focus on the resonant two-photon interaction. 
We leave detailed study of other interactions for future work.

\section{Two-photon coupling rate} \label{secTwoPhoton}

In this contribution, we only take into account the two-photon interaction that comes directly from the third-order coupling term $\propto \varphi_\res^2 \varphi_\atom$.
Depending on the working point, second-order terms in the expansion may also be non-zero, providing the linear coupling.
First, we illustrate how the two-photon coupling arises in the simplest case of negligible linear coupling.
Then, we investigate the case when the linear coupling is highly non-resonant.
Finally, we mention how the linear coupling can be suppressed.

\begin{figure}
\includegraphics{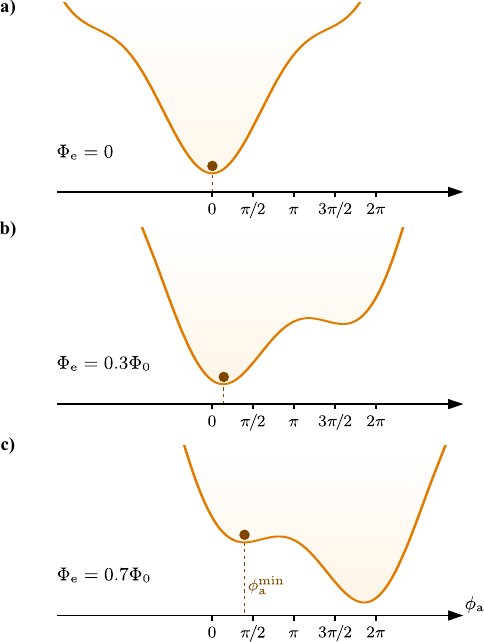}
\caption{Preparation of the rf SQUID metastable state using its magnetic bias~(see e.g.\ Ref.~\cite{poletto2009}).
a)~With a zero bias, the rf SQUID phase particle (filled circle) resides at $\phi_\atom^\mi = 0$ in equilibrium.
b)~Gradual sweep of the magnetic bias shifts the equilibrium.
c)~If needed, the system can be swept into a metastable position.
This configuration can be used for photodetection~\cite{opremcak2018measurement,opremcak2020high,shnyrkov2023rfsquid,stolyarov2023pnr,ilinskaya2024fluxqubit,sokolov2020superconducting}.
As in the text, $\phi_\atom$ is the rf SQUID (atom) phase and $\Phie$ is its biasing flux.
The rf SQUID inverse screening parameter is $\beta^{-1}_L = E^\atom_L/E^\atom_\JJ = 0.3$.
}
\label{figMetastablePreparation}
\end{figure}

\paragraph{Vanishing linear coupling.}
Assume for a moment that the capacitive coupling is negligible.
Consider the equilibrium phase difference over the coupling SQUID $\delta \approx -\pi/2$.
That can be accomplished by tuning the coupler external flux $\Phi_\cpl$ as discussed in what follows.
Then, according to the definition~\eqref{eqCouplerPhase} of $\delta$, leading terms in the coupling potential energy~\eqref{eqCouplerPotentialViaDepartures} are quadratic in either of the system variables:
\begin{equation}
\label{eqTwoPhotonEnergy}
	U_{c2} = -\hbar g_2 (a^\dag + a)^2 (b^\dag + b),
\end{equation}
and a similar one $\propto (a^\dag + a)(b^\dag + b)^2$.
For the two-photon resonance, the first contribution as in Eq.~\eqref{eqTwoPhotonEnergy}, predicts conversion of two photons in the resonator into a single excitation in the artificial atom and vice versa.
In Eq.~\eqref{eqTwoPhotonEnergy}, we denote the two-photon coupling rate by
\begin{equation}
\label{eqTwoPhotonRateMax}
	g_2 = \frac{E^\cpl_\JJ}{2\hbar} \varphi_{\res,\zpf}^2 \varphi_{\atom,\zpf},
\end{equation}
where the right-hand side is according to the expansion of Eq.~\eqref{eqCouplerPotentialViaDepartures}.

Note that the two-photon resonance occurs~\footnote{See, for example, Complement $\mathrm{C_{XX}}$ in the textbook~\cite{cohentannoudji2019quantum}, where a similar resonance conditions are obtained.} when the detuning $|\tilde\omega_\atom - 2\tilde\omega_\res|$ is either within the system linewidths or the two-photon coupling $g_2$.
Here $\tilde\omega_\res$ and $\tilde\omega_\atom$ stand for the observable (renormalized) frequencies of the resonator and the artificial atom.

Also note, that the two-photon coupling dominates over the capacitive single-photon one when 
\begin{equation}
\label{eqSmallCapacitiveCoupling}
	2E^\cpl_C n_{\res,\zpf} n_{\atom,\zpf}
		\ll E^\cpl_\JJ \varphi_{\res,\zpf}^2 \varphi_{\atom,\zpf}
			\sqrt{N_\mathrm{ch}},
\end{equation}
according to the Hamiltonian~\eqref{eq:ham_cpl} and expansion of the Hamiltonian~\eqref{eqCouplerPotentialViaDepartures}.
Here $N_\mathrm{ch}$ is the characteristic number of photons in the resonator.
That is, we compare the transition matrix elements $\bra{N_\mathrm{ch}{-1}} E_C^\cpl n_\atom n_\res \ket{N_\mathrm{ch}}$ and $\bra{N_\mathrm{ch}{-2}} U_\cpl \ket{N_\mathrm{ch}}$, where $\ket{N_\mathrm{ch}}$ is the highest resonator number state whose population is not negligible.

We assume that the rf SQUID is prepared in its metastable state (see Fig.~\ref{figMetastablePreparation}), providing the majority of the required shift $\delta$ in the equilibrium phases.
For simplicity, we assume that the coupler Josephson energy is small,
\begin{equation}
\label{eqSmallEjc}
	E^\cpl_\JJ \ll E_L^\res, \tilde E_L^\atom.
\end{equation}
In that case, it does not influence the equilibrium phases of both systems.
Right away, we determine $\phi_\res^\mi \approx 0$ according to Eqs.~\eqref{eqSmallEjc} and \eqref{eqResonatorExtremum}.

To determine the atom equilibrium and the two-photon coupling strength, we need to solve Eq.~\eqref{eqAtomExtremum} numerically.
For the rf SQUID bias at $\Phie = 0.7 \Phi_0$ and its inverse screening parameter $\beta_L^{-1} = E^\atom_L/E^\atom_\JJ = 0.3$, the rf SQUID equilibrium phase is $\phi^\mi_\atom \approx 0.396\pi$.
To arrive at phase difference $\delta$ at the coupler as required above, one biases it with $\Phi_\cpl \approx 0.1 \Phi_0$.
For the coupler Josephson energy $2E^\cpl_{1\JJ}/h = 10\,\mathrm{GHz}$ and magnitudes of zero-point fluctuations $\varphi_{\res,\zpf} = \varphi_{\atom,\zpf} \approx 0.156$, we estimate the two-photon coupling strength as $g_2 \approx 2\pi \times 18\,\mathrm{MHz}$.

Our choice of the magnitude of zero-point fluctuations corresponds to the 50\,Ohm impedance of the resonator.
The frequencies of the resonator and atom and hence the concrete circuit parameters are left almost arbitrary.
To clearly observe the two-photon coupling, it is only required that the resonator is in the two-photon resonance with the artificial atom.

According to Eq.~\eqref{eqTwoPhotonRateMax}, the two-photon coupling strength is proportional to the coupler Josephson energy $2E_{1\JJ}^\cpl$.
For a fixed distance between the contacts in the coupler Josephson junctions, an increase of $E_{1\JJ}^\cpl$ proportionally increases each junction internal capacitance and hence the coupler capacitive energy $E^\cpl_C$ according to Eq.~\eqref{eqCouplerCapacitiveEnergy}.
As $2E_{1\JJ}^\cpl / E^\cpl_C$ then stays the same, so does $g_2 / g_c$ the ratio of the two-photon coupling rate and the capacitive single-photon coupling rate.
Moreover, when the phases are well-defined variables according to the condition~\eqref{eqPhaseIsGoodQuantumNumber}, zero-point fluctuations in the charge variables $n_{\res,\zpf}$ and $n_{\atom,\zpf}$ can exceed unity, according to Eqs.~\eqref{eqResZPF} and \eqref{eqAtomZPF}.
That can make the condition~\eqref{eqSmallCapacitiveCoupling} of negligible capacitive coupling harder to hold, especially if the distance between the junction contacts is technologically fixed.
Therefore, we expect that the capacitive coupling often cannot be neglected.

\begin{figure}
\includegraphics{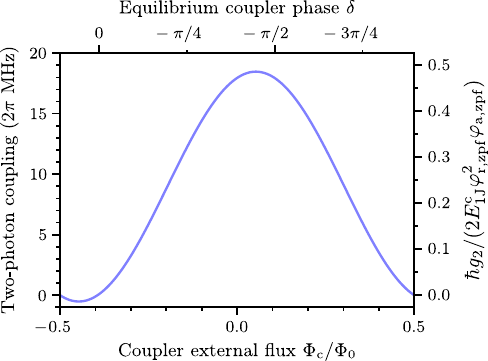}
\caption{Estimate for the two-photon coupling rate as a function of the coupler external flux (bottom axis) and the equilibrium phase difference at the coupler (top axis).
We do not take into account renormalization due to non-resonant interactions.
In the dimensionless axis, we use $2E^\cpl_{1\JJ}$ the Josephson energy of the coupler and $\varphi_{\res,\zpf}$ and $\varphi_{\atom,\zpf}$ the magnitudes of the zero-point fluctuations in the resonator and the artificial atom.
For the dimensionful axis, their values are as given in the text.
}
\label{figTwoPhotonRate}
\end{figure}

\paragraph{Non-resonant linear coupling.}
However, for a single mode resonator---such as an $LC$ tank---the single-photon coupling of any origin is highly off-resonant.
Moreover, in this contribution, we assume that it does not renormalize the two-photon coupling strength substantially.
For that case, we plot the dominating two-photon coupling strength
\begin{equation}
\label{eqTwoPhotonRate}
	g_2 = -\frac{E^\cpl_\JJ}{2\hbar}
		\varphi_{\res,\zpf}^2 \varphi_{\atom,\zpf} \sin\delta,
\end{equation}
in Fig.~\ref{figTwoPhotonRate} as a function of the coupler bias.
Equation~\eqref{eqTwoPhotonRate} provides the general expression for the coupling rate.

While Eqs.~\eqref{eqTwoPhotonRate} and \eqref{eqTwoPhotonEnergy} can be deduced from the expressions in Ref.~\cite{geller2015tunable} for a similar setup, two-photon coupling was not studied there.
Two-photon processes according to Eq.~\eqref{eqTwoPhotonEnergy} between lumped-element circuits were speculated about in Ref.~\cite{khabipov2022noncentrosymmetric} for a \emph{coupling rf SQUID}.
This reference also provides (see Appendix B therein) an analogue of Eq.~\eqref{eqTwoPhotonRateMax} for the interaction between resonator modes.
Equation~\eqref{eqTwoPhotonRateMax} is a partial case of Eq.~\eqref{eqTwoPhotonRate} with $\delta = 0$.

Note that for $\Phi_\cpl \approx \Phi_0/2$ and its multiples, the two-photon coupling switches off, according to Eq.~\eqref{eqTwoPhotonRate}.
In fact, according to Eqs.~\eqref{eqCouplerJosephsonEnergy} and \eqref{eqCouplerPotential}, all interactions of inductive type vanish.
The charging interactions persist, owing to the internal and possible shunting capacitances in the coupler.

\paragraph{Suppression of the linear coupling.}
We also mention, that the linear coupling can be suppressed by choosing the proper working point of the coupler.
In that point, its capacitive and inductive components cancel each other~\cite{neumeier2013single,ciani2019hamiltoncompcpbcrosskerr}.
We leave the detailed discussion of this regime for future work.

\section{Discussion}
\label{secDiscussion}

\paragraph{Bosonic approximation.}
We have expressed the atom variables in terms of bosonic ladder operators in Sec.~\ref{secQuantumHamiltonian}.
On the other hand, we assume the artificial to reside in its metastable state (see Fig.~\ref{figMetastablePreparation}c).
Hence, formally, the use of bosonic operators implies that the atom can be excited to the arbitrarily high energy states without leaving the metastable well. 
Moreover, the lowest energy state of the well is considered as the ground state of the atom.
Practically, using bosonic operators is an approximation only valid if we are not interested in the processes further away from the bottom of the metastable well.
It clearly breaks if the phase particle is excited high enough~\cite{poletto2009,opremcak2018measurement,opremcak2020high,shnyrkov2023rfsquid,stolyarov2023pnr,ilinskaya2024fluxqubit,sokolov2020superconducting} to enter the main well containing the true ground state of the atom.
However, our main analytical result for the two-photon coupling rate can be easily generalized beyond the bosonic approximation as in Sec.~\ref{secApplications}.
In fact, in this contribution, we have used that approximation mainly for brevity of presentation and to connect to future work.

\paragraph{Validity of the cosine expansion.}
To expand the coupler Josephson energy in Eq.~\eqref{eqCouplerPotential}, we have assumed that departures from the equilibrium phases are small.
We now provide concrete criteria for that.
One can require that $\varphi_{\res,\zpf} N_\mathrm{ch} \ll 1$ and $\varphi_{\atom,\zpf} \ll 1$, where we assume that the artificial atom cannot be excited beyond its first excited state.
When these criteria hold, we expect that all higher expansion terms in the Josephson energy are negligible.
However, the two-photon coupling rate also becomes small, compared to the possible inductive coupling strength.

In fact, for our estimates in the previous section, the criterion above does not hold when there is more than one photon in the resonator.
However, we claim that it is enough to satisfy a weaker condition,
\begin{equation}
\label{eqSmallZPFs}
	\varphi_{\res,\zpf}^2 \ll 1,
\quad
	\varphi_{\atom,\zpf}^2 \ll 1,
\end{equation}
for Eq.~\eqref{eqTwoPhotonRate} for the two-photon rate to be approximately correct.
For example, consider a term with $\varphi_{\res,\zpf}^2 \varphi_{\atom,\zpf}^3 a^2 b^{\dag2} b$ in the expansion of the coupler Josephson energy.
This and other similar terms give rise to two-photon processes as well.
However, under the condition above, such processes occur rarely compared to the processes with the rate as in Eq.~\eqref{eqTwoPhotonRate}.
To neglect the processes of higher-order in the resonator operators, one also needs to make sure that the photon number is small enough.
Note that the conditions in Eqs.~\eqref{eqSmallZPFs} follow from the definitions in Eqs.~\eqref{eqResZPF}--\eqref{eqAtomZPF} and the condition~\eqref{eqPhaseIsGoodQuantumNumber}.

\paragraph{Capacitive coupling.}
While we have assumed that the linear coupling is non-resonant in Sec.~\ref{secTwoPhoton}, it is still worth comparing the two-photon coupling to the capacitive coupling that arises due to the junction charging energy.
To achieve the coupler Josephson energy as in the previous section, each of its junctions should have the critical current of $I_0^\cpl \approx 10.1$\,nA.
Assuming the critical current density in a Josephson junction of $1.7\,\mathrm{\mu A/\mu m^2}$ and its capacitance density of $50\,\mathrm{fF/\mu m^2}$~\cite{steff2006, lisenfeld2007thesis}, we estimate the coupler self-capacitance to be about $2C_{1\JJ}^\cpl \approx 0.6\,\mathrm{fF}$ for the above critical current.
Note that the coupler comprises two junctions in parallel.

According to Eqs.~\eqref{eq:ham_cpl} and \eqref{eqCouplerCapacitiveEnergy}, capacitive coupling rate reads, for $C'_\cpl \ll C_\res, C_\atom$:
\begin{equation}
	g_c \approx \frac{4e^2 C'_\cpl}{\hbar C_\res C_\atom} n_{\res,\zpf} n_{\atom,\zpf}.
\end{equation}
We estimate $g_c$ solely due to the internal capacitances, i.e.\ when the coupler shunting capacitance $C_\cpl = 0$ and $C'_\cpl = 2C_{1\JJ}^\cpl$.
In line with the previous estimates, $n_{\res,\zpf} = n_{\atom,\zpf} = 1/2\varphi_{\res,\zpf} = 1/2\varphi_{\atom,\zpf} \approx 3.205$.
We assume that $10 E^\cpl_\JJ = E^\res_L$ accordingly with the criterion in Eqs.~\eqref{eqSmallEjc}.
Hence the resonator plasma frequency $\omega^p_\res = \sqrt{8E^\res_C E^\res_L}/\hbar = 2\varphi_{\res,\zpf}^2 E^\res_L / \hbar$ should be at least about $2\pi \times 4.9$\,GHz for our choice of $\varphi_{\res,\zpf}$ and $E_\JJ^\cpl \leq 2 E_{1\JJ}^\cpl$.
From above, we bound $\omega^p_\res$ to be up to $2\pi \times 10$\,GHz, so that it is much smaller than the gap~\cite{court2008energygap,kittel2005intro} $\Delta \sim 100$\,GHz of superconducting aluminum.
For this frequency range and $\varphi_{\res,\zpf}$, the resonator capacitance $C_\res$ should be approximately from 320 to 650\,fF.
To approximately match the two-photon resonance, we assume that the atom plasma frequency $\omega^p_\atom = \sqrt{8E^\atom_C\tilde{E}^\atom_L}/\hbar \approx 2\omega^p_\res$.
This yields $C_\atom \approx C_\res/2$ as we set $\varphi_{\res,\zpf} = \varphi_{\atom,\zpf}$ and $\tilde E_L^\atom / E_C^\atom = E_L^\res / E_C^\res$.
With the above assumptions, one obtains the capacitive coupling rate $g_c/2\pi$ from 4.5 to 20\,MHz.

\paragraph{Limits on the two-photon coupling strength.}
In our estimates, the magnitude of the two-photon coupling $g_2$ is limited by our assumption in Eqs.~\eqref{eqSmallEjc} that the coupler does not influence the equilibria of the coupled system.
This requirement can be lifted for a weaker one: that the coupler energy does not yield ultrastrong~\cite{forndiaz2019ultrastrong,kockum2019ultrastrong} coupling.
In that case, we also expect that the capacitive coupling can be of lower rate in the estimate above;
accordingly, the two-photon coupling with the rate of tens of megahertz can dominate even if the single-photon coupling is not off-resonant.

\paragraph{Effect of inductance of the big loop.}
In the big loop in Fig.~\ref{figScheme}, we only take into account the resonator inductance.
However, other wires in the loop may provide significant geometrical and kinetic inductance.
Additional inductance $L'$ is taken into account in Eq.~\eqref{eqBigLoopFullFlux} by changing $L_\res \to L_\res + L'$.
Consequently, the phase drop across the resonator capacitance becomes only a part of the full flux $\Phi_\Sigma$ in the respective units, i.e. $\phi_\res \to \lambda \phi_\res$ with $\lambda = L_\res / (L_\res + L')$.
In the resonator Lagrangian (see the \hyperref[apCircuitHamiltonian]{Appendix}), that is equivalent to reduction of the resonator capacitance $C_\res \to \lambda^2 C_\res$.
Subsequently, that increases in the resonator frequency and vacuum fluctuations of its phase.
Moreover, we need to replace $\phi_\res$ with the reduced one in the coupler Josephson energy in Eq.~\eqref{eqCouplerJosephsonEnergy}.
The net effect is the reduction in the inductive couplings, $g_i \to \sqrt\lambda g_i$ and $g_2 \to \lambda g_2$.
Here we have introduced the single-photon inductive coupling rate $g_i = (E^\cpl_\JJ / \hbar) \varphi_{\res,\zpf} \varphi_{\atom,\zpf} \cos\delta$ according to expansion of Eq.~\eqref{eqCouplerPotentialViaDepartures}.
Kinetic inductance can be taken into account in the same way.

\paragraph{Vortex-induced resistivity.}
Possible fluctuations of superconducting condensate phases average the two-photon coupling.
According to Fig.~\ref{figTwoPhotonRate}, that reduces the coupling magnitude when close to its highest value.
We expect that the resistive conduction of Cooper pairs may give rise to such fluctuations.
It is known, that movement of vortices results in such resistivity~\cite{tinkham2004intro,song2009responsevortices}.
We have discussed how they can be avoided in Sec.~\ref{secCurrentFreePath}.

\section{Possible applications}
\label{secApplications}

\paragraph{Photodetection.}
We have shown a sizeable two-photon coupling of a resonator to an rf SQUID.
First of all, it is tempting to study its use for detection of microwave photon pairs.
In other words, the rf SQUID may act as a Josephson photomultiplier~\cite{opremcak2018measurement,opremcak2020high,shnyrkov2023rfsquid,stolyarov2023pnr,ilinskaya2024fluxqubit} with a partial photon-number resolution~\cite{sokolov2020superconducting} and enhanced detection speed owing to non-perturbative two-photon coupling.

\paragraph{Other systems to couple.}
The theory in the present paper can be used to engineer two-photon coupling between a resonator and some other types of artificial atoms.
Transmon~\cite{koch2007transmon,blais2021circuit} states center at the zero phase difference across its junction, i.e. $\phi^\mi_\atom = 0$ for it.
Thus, a transmon cannot break the Josephson coupling symmetry when no coupler bias is applied.
With proper equilibrium phases, Eqs.~\eqref{eqTwoPhotonRate} and \eqref{eqCouplerPhase} can be used to determine the transmon coupling rate.
However, Josephson coupling energy $E_\JJ^\cpl$ in Eq.~\eqref{eqCouplerJosephsonEnergy} vanishes at $\pi\Phic/\Phi_0 = \pi/2$ where the rest of the expression in Eq.~\eqref{eqTwoPhotonRate} reaches its maximal absolute value.
A simple calculation shows that the coupling highest magnitude is approximately half as small as for the phase qubit in Sec.~\ref{secTwoPhoton}.
Asymmetrical coupling SQUID as in Ref.~\cite{ayyash2024driven} allows for better figures;
alternatively, one can bias the big loop in Fig.~\ref{figScheme} instead of the coupling SQUID, similarly to Fig.~\ref{figSymmetry}.

In the flux qubit~\cite{orlando1999threejjqubit,mooij1999josephson}, energy eigenstates are shifted in the phase difference on the qubit junctions. 
However, a correcting coupler bias may still be required.
Moreover, the flux qubit energy states are highly anharmonic.
Hence Eq.~\eqref{eqTwoPhotonRate} for the two-photon coupling strength modifies to
\begin{equation}
	g_2 = -\frac{E^\cpl_\JJ}{2\hbar} \bra1 \hat\varphi_\atom \ket0
		\varphi_{\res,\zpf}^2 \sin\delta,
\end{equation}
where $\ket0$ and $\ket1$ are the first two energy states.

\paragraph{Direct waveguide coupling.}
A dc SQUID coupled to a waveguide~\cite{wustmann2013parametric,petrovnin2023microwave} may be used to achieve two-photon coupling between the waveguide modes, similarly to coupling of two separate circuits as we have discussed.
However, directly attaching a waveguide breaks the superconducting loop that ties the system static phases.
Recall that this tie is what allows for breaking the Josephson symmetry and makes the two-photon interaction possible.
To overcome this problem, the waveguide can be shunted with a large inductance.
While the signal currents will avoid flowing through it, such an inductance allows the dc screening currents to pass.

\paragraph{Second harmonic generation.}
A dc SQUID at the end of a waveguide may be used for second harmonic generation, providing a two-photon coupling between its modes.
This scheme looks especially appealing, as it does not use capture cavities or artificial atoms to mediate the excitations.
That avoids non-radiative processes in them, as well as the scattering losses due to the limited bandwidth of these systems.
Hence, it may be possible to manipulate the carrier frequency of extremely short pulses of arbitrary shape arriving at arbitrary times.

One possible limit for the scheme bandwidth is the onset of linear interaction.
When coupling to the continuum of modes, linear interaction with some of them is always resonant.
Hence it must be suppressed as outlined above and in Refs.~\cite{neumeier2013single,ciani2019hamiltoncompcpbcrosskerr}.
However, capacitive and inductive waveguide couplings depend differently on the frequency~%
\footnote{Introducing the mode operators so that the respective waveguide eigenmodes respect the boundary condition in a reasonable way allows one to avoid direct interaction between them (see Ref.~\cite{sokolov2023thesis} and the references therein).
That yields the coupling rates that depend on the frequency in the same manner for the capacitive and the relevant type of inductive coupling~\cite{shitara2021nonclassicality}.
The case when the same semi-infinite waveguide is coupled both capacitively and inductively was considered in Ref.~\cite{parrarodriguez2018quantum}.
In that case, the coupling rates are of different functional form.
Note that in all mentioned references, the waveguide modes are assumed to be (approximately) TEM.
}.
With considerable detuning, they will cease to cancel each other.

Second harmonic generation in a related scheme was observed in Ref.~\cite{khabipov2022noncentrosymmetric} using an rf SQUID in the resonator.
Note that, second harmonic generation works without a pump tone.
That is unlike frequency conversion with a pumped dc SQUID in a superconducting resonator in Refs.~\cite{zakkabajjani2011quantum,lu2023resolvingfock} or with a two-photon process that again involves a pump to interface superconducting qubit with Rydberg atoms as mentioned in Refs.~\cite{kurizki2015quantum,xiang2013hybrid}.

\section{Conclusions}
\label{secConclusions}

To conclude, we have shown that a two-photon interaction can be mediated by a symmetric dc SQUID.
We have estimated its magnitude and studied its dependence on the coupler magnetic field.
In addition to coupling of the resonator and phase qubit, we have outlined several other schemes relevant for applications.

Two-photon interaction arises owing to two related mechanisms:
In one of them, the external magnetic field in the coupling SQUID breaks the Josephson symmetry.
In the other, one of the interacting parties provides an additional asymmetry through the equilibrium drop of the condensate phase on it.
The latter is the case for the metastable state of the rf SQUID that we consider.
We have shown that the magnetic-field symmetry breaking can be re-formulated as the mechanism due to non-zero equilibrium phases.
On the other hand, an rf SQUID state with a non-zero equilibrium phase is typically prepared using its magnetic field bias.

We have estimated the rate of two-photon coupling that arises due to both mechanisms.
We expect that the rate up to 18\,MHz is attainable, which is comparable to our estimate of the capacitive coupling strength in the circuit.
We have argued that the coupling rate can be even higher and outlined some possible uses of the predicted interaction.
We have shown that the two-photon interaction can be switched off gradually by changing the coupler bias.

Symmetry breaking due to a non-zero equilibrium phase allows one to maximize the two-photon coupling via the symmetrical coupling SQUID.
Apart from the phase qubit case that we consider, we expect this symmetry-breaking mechanism to contribute to the flux qubit coupling.
On contrary, that is not the case for the transmon qubit, for which we have outlined other approaches.

\begin{acknowledgments}

E.~V.~S.\ acknowledges support from the National Research Foundation of Ukraine through the project No.\ 2023.03/0165, Quantum correlations of electromagnetic radiation.
A.~M.~S.\ acknowledges partial support from the Academy of Finland (Contract No.\ 321982) and the President's scholarship for young scientists;
he also thanks T.~T.~Heikkil\"a for pointing out a similarity with the superconducting diode effect.
We are indebted to A.~Parra-Rodriguez for enlightening comment about the frequency dependence of waveguide couplings.
We thank M.~Ayyash for pointing out a closely related paper which he co-authored, as well as some other related works.
\end{acknowledgments}

\appendix

\section{Derivation of the circuit Hamiltonian}
\label{apCircuitHamiltonian}

\begin{figure}
	\centering
	\includegraphics{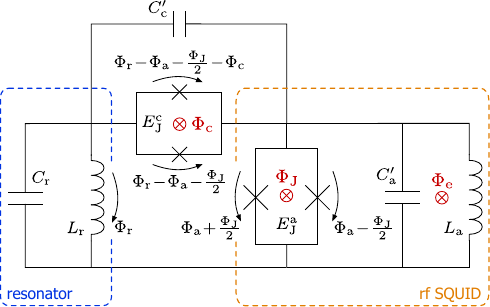}
	\caption{Circuit diagram of an rf SQUID with variable Josephson energy, coupled to the resonator.
	Their interaction is via the coupling dc SQUID and a capacitance.
	Arrows indicate the direction from higher to lower node flux.}
	\label{figDetailedCircuit}
\end{figure}

Here we derive the Hamiltonian of the circuit in Fig.~\ref{figDetailedCircuit}.
Instead of a single Josephson junction in the rf SQUID loop in Fig.~\ref{figScheme}, we consider a dc SQUID there.
It can be useful for the applications where the rf SQUID resonance should be changed \emph{in situ}.
The case of the circuit with a dc SQUID contains the single-junction one as a partial case.

The Lagrangian of the circuit is
\begin{equation} \label{eq:lgr}
	L = T - U.
\end{equation}
Its kinetic energy consists of the charging energies of the circuit parts:
\begin{equation} \label{eq:kin}
	T = \frac{C_\res \dot{\Phi}_\res^2}{2}
		+ \frac{C'_\cpl (\dot{\Phi}_\res - \dot{\Phi}_\atom)^2}{2}
		+ \frac{C'_\atom \dot{\Phi}_\atom^2}{2}.
\end{equation}
By $C'_\cpl = C_\cpl + 2 C_{1\JJ}^\cpl$ we denote the total capacitance of the coupler.
Note that since we consider a dc SQUID instead of a single junction, the total capacitance of the rf SQUID is now $C'_\atom = C_\atom + 2C^\atom_{1\JJ}$.
Each total capacitance consists of the shunt capacitance and two capacitances of the respective Josephson junctions.
$\dot\Phi_\res$ and $\dot\Phi_\atom$ are the voltages at the resonator and the rf SQUID.
We have used the Kirchhoff's contour rule for the voltages to express the voltage drop on the coupler.
Potential energy of the full circuit reads
\begin{equation} \label{eq:U1}
	U = U_\res + U_\atom + U_\cpl.
\end{equation}
Consider now the potential energies of each circuit part.

For the resonator, it is solely of inductive origin:
\begin{equation}
\label{eq:Ures}
	U_\res = \frac{\Phi_\res^2}{2L_\res},
\end{equation}
where $\Phi_\res$ is the resonator node flux~\cite{devoret1997}.
On the other hand, the rf SQUID potential energy
\begin{multline}
\label{eqAtomPotential1}
	U_\atom = \frac{(\Phi_\atom - \Phie - \PhiJ/2)^2}{2L_\atom}
		- E^\atom_{1\JJ} \cos \left(2\pi\frac{\Phi_\atom + \PhiJ/2}{\Phi_0}\right)
\\
		- E^\atom_{1\JJ} \cos \left(2\pi\frac{\Phi_\atom - \PhiJ/2}{\Phi_0}\right)
\end{multline}
has an inductive and a Josephson part.

First consider the Josephson part in the last two terms.
In the rf SQUID, $\Phi_\atom + \Phi_\JJ/2$ is the drop of the node flux on the left junction in its SQUID, as shown in Fig.~\ref{figDetailedCircuit}.
In other words, it is the difference in the superconducting condensate phase over the left junction, up to a constant factor.
With the drop of $\Phi_\atom - \Phi_\JJ/2$ over the right junction, the total flux in the SQUID loop is $\Phi_\JJ$~\cite{feynman1965feynmanlectures}.
That is a result of the flux quantization rule~\cite{vool2017,barone1982physicsappsjj,orlando1999threejjqubit}.
Its origin and applicability are discussed in Sec.~\ref{secClassicalHamiltonian}.

The first term in Eq.~\eqref{eqAtomPotential1} is the rf SQUID energy due to the self-inductance of its loop.
The loop current induces a flux which is a difference between the full flux and the externally imposed flux $\Phie$.
The full flux in the rf-SQUID loop is equal to the node flux drop over the right junction in its dc SQUID, according to the contour rule mentioned above.
Hence the argument in the inductive energy in Eq.~\eqref{eqAtomPotential1}.

The coupler potential energy is of Josephson origin:
\begin{multline}
\label{eqCouplerPotential1}
	U_\cpl =
		- E^\cpl_{1\JJ} \cos\left(2\pi\frac{\Phi_\res - \Phi_\atom
										 - \PhiJ/2 - \Phic}
									  {\Phi_0}\right)
\\
		- E^\cpl_{1\JJ} \cos\left(2\pi\frac{\Phi_\res - \Phi_\atom - \PhiJ/2}
										{\Phi_0}\right).
\end{multline}
Cosine arguments in the equation are the junction phase differences in the top and the bottom junctions, as shown in Fig.~\ref{figDetailedCircuit}.
Such phases compensate the external flux over the loop.
One can also check, that then no flux threads the ``inner'' loop in the figure---formed by the resonator inductance, lower junction of the coupler SQUID, and the left junction of the dc SQUID inside the rf SQUID.

The generalized momenta $Q_\res$ and $Q_\atom$ are given by
 \begin{equation}  \label{eq:Qs}
 	\begin{split}
 		Q_\res & = \frac{\partial L}{\partial \dot{\Phi}_\res} = - C'_\cpl \dot{\Phi}_\atom + (C_\res + C'_\cpl) \dot{\Phi}_\res, \\
  		Q_\atom & = \frac{\partial L}{\partial \dot{\Phi}_\atom} = (C'_\atom + C'_\cpl) \dot{\Phi}_\atom - C'_\cpl \dot{\Phi}_\res,
 	\end{split}
 \end{equation}
i.e., they are the charges accumulated on the capacitances of the resonator and the rf SQUID.
The Hamiltonian is obtained via the Legendre transform:
\begin{align}
\label{eq:lgndr}
	\Ham &= Q_\res \dot{\Phi}_\res + Q_\atom \dot{\Phi}_\atom - L
\\
\label{eq:ham0}
		 &= \frac{Q^2_\res}{2\tilde{C}_\res}
		   + \frac{Q_\atom^2}{2\tilde{C}_\atom}
		   + \frac{\tilde{C}_\cpl}{C'_\atom C_\res} Q_\atom Q_\res
	 	   + U,
\end{align}
where the capacitances $\tilde{C}_\res$, $\tilde{C}_\atom$, and $\tilde{C}_\cpl$ are given by
\begin{gather}
\label{eqCaCr}
		\tilde{C}_\res = C_\res + \frac{C'_\atom C'_\cpl}{C'_\atom + C'_\cpl},
\quad
		\tilde{C}_\atom = C'_\atom + \frac{C_\res C'_\cpl}{C_\res + C'_\cpl},
\\
\label{eqCcpl}
	\tilde{C}_\cpl = \left(\frac{1}{C'_\atom} + \frac{1}{C_\res} + \frac{1}{C'_\cpl}\right)^{-1}.
\end{gather}
To avoid lengthy algebra in obtaining Hamiltonian~\eqref{eq:ham0}, one can use the trick from Appendix~A of Ref.~\cite{sokolov2020superconducting}.

The renormalized capacitances in Eqs.~\eqref{eqCaCr}--\eqref{eqCcpl} can be easily interpreted.
$\tilde C_\cpl$ is the coupling capacitance loaded by the series connection with the resonator capacitance $C_\res$ and the total capacitance $C'_\atom$ of the atom.
$\tilde C_\res$ is the resonator capacitance loaded by the parallel connection with the full atom capacitance $C'_\atom$ that is connected in series with the full coupling capacitance $C'_\cpl$.
The expression for $\tilde C_\atom$ is interpreted analogously.

We express each sum of cosines as a product in Eqs.~\eqref{eqAtomPotential1} and \eqref{eqCouplerPotential1}.
The rewritten potentials of the rf SQUID and the coupler are
\begin{align}
\label{eqAtomPotential2}
	&U_\atom = \frac{(\Phi_\atom - \Phie - \PhiJ/2)^2}{2L_\atom}
		- 2E^\atom_{1\JJ} \cos \frac{\pi\PhiJ}{\Phi_0}
			  			 \cos \frac{2\pi\Phi}{\Phi_0},
\\
\label{eqCouplerPotential2}
	&U_\cpl = -2E^\cpl_{1\JJ} \cos \frac{\pi\Phic}{\Phi_0}
		  \cos \Big(2\pi \frac{\Phi_\res - \Phi_\atom - \Phic/2 - \PhiJ/2}
							{\Phi_0}\Big).
\end{align}	
That is, SQUID potential energy is similar to the energy of a single junction with a Josephson energy set by the external flux.

We recover the case of a single-junction rf SQUID in Fig.~\ref{figScheme}.
For that, we set $\Phi_\JJ = 0$ in the potential energies above.
We use that two junctions in parallel are equivalent to a single one with twice the capacitance, $C^\atom_\JJ = 2C^\atom_{1\JJ}$, and twice the critical current and Josephson energy, $E^\atom_\JJ = 2E^\atom_{1\JJ}$.
We also define the Cooper-pair numbers $n_\atom=Q_\atom/2e$ and $n_\res=Q_\res/2e$ and the phases $\phi_\atom=2\pi\Phi_\atom/\Phi_0$ and $\phi_\res=2\pi\Phi_\res/\Phi_0$ in Eqs.~\eqref{eq:ham0} and \eqref{eqAtomPotential2}--\eqref{eqCouplerPotential2}.
With that, we arrive at the circuit Hamiltonian in Eqs.~\eqref{eqClassicalHamiltonian}--\eqref{eqCouplerPotential}.

\bibliography{bibliography,common_sources,counters,jj_sources,sokolov,cqed,nonlin,optmech,param,circulators,gates}

\end{document}